\documentclass[aps,pre,showpacs,twocolumn,superscriptaddress,amssymb]{revtex4}
\usepackage{graphicx}
\usepackage{txfonts}

\usepackage{amssymb}

\usepackage{makeidx}

\def\simge{\mathrel{%
    \rlap{\raise 0.511ex \hbox{$>$}}{\lower 0.511ex \hbox{$\sim$}}}}
\def\simle{\mathrel{
    \rlap{\raise 0.511ex \hbox{$<$}}{\lower 0.511ex \hbox{$\sim$}}}}

\newcommand \beq{\begin{eqnarray}}
\newcommand \eeq{\end{eqnarray}}
\newcommand{\del}{\partial}

\begin{document}

\title{Universal shocks in random matrix theory}

\author{Jean-Paul Blaizot}
\email{Jean-Paul.Blaizot@cea.fr} \affiliation{IPTh, CEA-Saclay,
91191 Fif-sur Yvette, France}

\author{Maciej A. Nowak}
\email{nowak@th.if.uj.edu.pl} \affiliation{M. Smoluchowski Institute
of Physics and Mark Kac Center for Complex Systems Research,
Jagiellonian University, PL--30--059 Cracow, Poland}

\date{\today}


\begin{abstract}
We link  the appearance of universal kernels in random matrix
ensembles to the phenomenon of  shock formation in some fluid dynamical equations. Such equations are derived from Dyson's random walks after a proper rescaling of the time. In the case of the Gaussian
Unitary Ensemble, on which we focus in this letter, we show that the orthogonal polynomials, and their Cauchy transforms, evolve
according to a viscid Burgers equation with an effective
``spectral  viscosity'' $\nu_s=1/2N$, where $N$ is the size of the
matrices. We relate the edge of the spectrum of eigenvalues to the shock that naturally appears in the Burgers equation for appropriate initial conditions, thereby obtaining a new perspective on universality.

\end{abstract}



\maketitle


In a seminal paper~\cite{DYSON}, Dyson showed that the distribution of eigenvalues of a random matrix could be interpreted as the result of a random walk performed independently by each of the matrix elements. The resulting distribution yields the so called ``Coulomb gas'' picture, with the eigenvalues identified to charged point particles repelling each other according to  Coulomb law. For matrices of large sizes, this   correctly describes the bulk
properties of the spectrum~\cite{WEIDEN}.
In his
original work Dyson introduced a restoring force preventing the
eigenvalues to spread for ever as time goes. This is what allowed him to find a stationary solution corresponding to the random ensemble considered  with a chosen variance
(related to the restoring force). In this note, we point out that a richer structure emerges if one ignores the restoring force and performs a rescaling of the time of the random walk. The random walk can then be described by an equation of fluid dynamics, the viscid Burgers equation. In this picture, that we may refer to as ``Dyson fluid'', the edge of the spectrum appears as the precursor of a shock wave, and its universal properties follow from a simple analysis of the Burgers equation that was developed in other contexts ~\cite{BESSIS}.  This enables us to recover familiar results of random matrix theory in a simple way, and get a new perspective on the issue of universality.

In this letter we shall consider only the Gaussian Unitary Ensemble, although we believe that many of our results can be extended to other ensembles. Thus, we consider
 $N\times N$ hermitian matrices $H$ with complex entries.
We assume that these matrices evolve in time according to the following random walk:
 in the  time step $\Delta\tau$ the matrix elements change
according to $ H_{ij}\to H_{ij}+\delta H_{ij}$, with $\langle \delta
H_{ij}\rangle=0$, and $\langle (\delta
H_{ij})^2\rangle=(1+\delta_{ij})\Delta t. $ That is , we assume that
at each time step, the increment of the matrix elements follows from a
Gaussian distribution with a variance proportional to $\Delta t$. The initial condition on the random walk is that at time $t=0$, all the matrix elements vanish.
Alternatively, let $x_i$ denote  the eigenvalues of $H$. The
previous random walk translates into a corresponding  random walk of
the eigenvalues,  with the following characteristics~\cite{DYSON} \beq \langle
\delta x_i \rangle =E(x_i)\, \Delta t, \qquad \langle( \delta x_i)^2
\rangle = \Delta  t, \eeq where the ``Coulomb force'' \beq
E(x_j)=\sum_{i\ne j}\left(  \frac{1}{x_j-x_i}\right)\eeq originates
in the Jacobian of the transformation from the matrix elements to
the eigenvalues, $\Delta=\prod_{i<j}(x_i-x_j)^2$. The probability for finding  the set of eigenvalues
$x_1,\cdots,x_N$ at time $t$ obeys the Smoluchowski-Fokker-Planck
equation \beq\label{SFP} \frac{\del P}{\del
t}=\frac{1}{2}\sum_i\frac{\del^2 P}{\del x_i^2}-\sum_i
\frac{\del}{\del x_i}\left( E(x_i)P\right), \eeq  whose solution reads \beq
P(x_1,\cdots,x_N,t)=C\prod_{i<j}(x_i-x_j)^2 \, {\rm
e}^{-\sum_i\frac{ x_i^2}{2t}}, \eeq with $C$ a (time-dependent)
normalization constant.

The average density of eigenvalues,  $\tilde \rho(x)$, may be
obtained from $P$ by integrating over $N-1$ variables. Specifically:
\beq \tilde \rho(x,t)= \int \prod_{k=1}^N dx_k
\,P(x_1,\cdots,x_N,t)\sum_{l=1}^N\delta(x-x_l), \eeq with
normalization  $ \int dx\, \tilde\rho(x)=N. $
Similarly we define the ``two-particle'' density $
\tilde\rho(x,y)=\langle\sum_{l=1}^N\sum_{j\ne
l}\delta(x-x_l)\delta(y-x_i)\rangle, $ with $ \int dx\, dy
\,\tilde\rho(x,y)=N(N-1). $ These various densities obey an infinite
hierarchy of equations  obtained form Eq.~(\ref{SFP})  for $P$. Thus, the equation relating the one and two particle
densities reads \beq \frac{\del\tilde\rho(x,t)}{\del
t}=\frac{1}{2}\frac{\del^2
\tilde\rho(x,t)}{\del\lambda^2}-\frac{\del}{\del\lambda} \fint dy
\,\frac{\tilde\rho(x,y,t)}{x-y}, \eeq where $\fint $ denotes the
principal value of the integral.

In  the large $N$ limit, this equation becomes a closed equation for the one particle density.
 To show that, we set
 \beq
\tilde\rho(x,y)=\tilde\rho(x)\tilde\rho(y)+\tilde\rho_{con}(x,y),
\eeq where $\tilde\rho_{con}(x,y)$ is the connected part of the
two-point density. Then we change the normalization of the single
particle density, defining \beq \tilde\rho(x)=N\rho(x), \eeq and
similarly $\tilde\rho(x,y)=N(N-1)\,\rho(x,y)$. At the same time, we
rescale the time so that $\tau=N t$~\cite{BS}. One then obtains \beq\label{eqrhoexact}
\frac{\del\rho(x)}{\del \tau}\!+\! \frac{\del}{\del x}\rho(x)\!
\fint\! dy\,\frac{\rho(y)}{x-y}=\frac{1}{2N}\frac{\del^2
\rho(x)}{\del x^2}\!+\! \fint dy \,\frac{\rho_{con}(x,y)}{x-y} .
\eeq In the large $N$ limit, the right hand side is negligible, leaving as announced a closed equation for $\rho(x,\tau)$. This equation can be further transformed into an equation for the  resolvent \beq
G(z,\tau)=\left< \frac{1}{N}{\rm Tr} \frac{1}{z-H(\tau)}\right>=\int
dy \,\frac{\rho(y,\tau)}{z-y}, \eeq whose imaginary part for
$z=x-i\epsilon$, and $x$ real, yields the average spectral density
$\rho(x)$, and the real part is the Hilbert transform of $\rho$, $
{\cal H}\rho(x)=\fint dy \,{ \rho(y)}/{(x-y)}$. By taking the
Hilbert transform of Eq.~(\ref{eqrhoexact}), keeping only the dominant term in the large $N$ limit and using
well known properties of the Hilbert transform, one gets  the following equation for $G(z,\tau)$ \beq\label{inviscBurgers}
\partial_{\tau} G(z, \tau)+ G(z,\tau)\,\partial_z G(z, \tau)= 0.
\eeq This  is the  inviscid Burgers equation~\cite{VOICULESCU}. Note
that the Laplacian, that naturally appears in the description of
diffusive processes, has  disappeared in the large $N$ limit, as
already indicated  after Eq.~(\ref{eqrhoexact}). The non-linear term
plays  the role of an effective mean field representing the mutual
repulsion of the evolving eigenvalues. We shall return later to the
role of diffusion, and focus now on the solution of
Eq.~(\ref{inviscBurgers}).

This can be obtained by using the method of (complex)
characteristics~\cite{BN1}, with the characteristics determined by
the implicit equation $z=\xi+\tau G_0(\xi)$, where
$G_0(z)=G(z,\tau=0)=1/z$. Assuming the solution $\xi(z,\tau)$ to be
known, the Burgers equation can be solved parameterically as
$G(z,\tau)=G_0(\xi(z,\tau))=G_0(z-\tau G(z,\tau))$. The solution of
this equation that is analytic in the lower half plane is
 \beq G(z,\tau)
=\frac{1}{2\tau}(z-\sqrt{z^2-4\tau})\label{semicircle}\, ,\eeq
whose imaginary part yields the familiar Wigner's semicircle for the average density of eigenvalues. In the fluid dynamical picture suggested by the Burgers equation, the edge of the spectrum corresponds to a singularity that is associated with the precursor of a shock wave, sometimes  referred to as a ``pre-shock'' ~\cite{BESSIS}. This singularity occurs when the mapping between $z$ and $\xi$ ceases to be one-to-one, a condition required for the validity of the method of characteristics.  This takes place when
$dz/d\xi=0=1+\tau G'_0(\xi_c)$, defining $\xi_c(\tau)$. Since $G_0'(\xi_c)=-1/\xi_c^2$, $\xi_c(\tau)=\pm \sqrt{\tau}$, and  $z_c=\xi_c+\tau G_0(\xi_c)=\pm 2\sqrt{\tau}$. That is, the singularity occurs precisely at
the edge of the spectrum. Furthermore, the resulting singularity is of the
square root type. To see that, let us expand the characteristic equation around the singular point. We get
\beq
z-z_c=\frac{\tau}{2}(\xi-\xi_c)^2 G_0''(\xi_c)=\frac{\tau}{\xi_c^3}(\xi-\xi_c)^2 .
\eeq
It follows that, in the vicinity of the positive edge of the spectrum $z\simeq z_c= 2\sqrt{\tau}$,  $\xi-\xi_c=\pm \tau^{1/4}\sqrt{z-z_c}$.  Thus, as $z$ moves towards $z_c$ and is bigger than $z_c$, $\xi$ moves to $\xi_c$ on the real axis. When $z$ becomes smaller than $z_c$, $\xi$ moves away from $\xi_c$ along the imaginary axis. The imaginary part therefore exists for $z<z_c$ and yields a spectral density $\rho(z)\sim \sqrt{z_c-z}$, in agreement with (\ref{semicircle}). This square root behaviour of the spectral density implies that in the vicinity of the edge of the spectrum, the number of eigenvalues in an interval of width $s$ scales as  $N s^{3/2}$, implying that the interlevel spacing
goes as $N^{-2/3}$.

In order to capture the fine structure of the level density in the vicinity of the edge, we need to take into account the $1/N$ corrections. We wish to do that within the Dyson fluid picture. Trying to go beyond the large $N$ limit at the level of Eq.~(\ref{eqrhoexact})  requires  in particular   handling   the connected two-point function. We shall not do so directly, but shall rely on time dependent
orthogonal polynomials, i.e., extend to our time-dependent setting tools that are familiar in random matrix theory. This will allow us to obtain simple equations that generalize Eq.~(\ref{eqrhoexact}) (albeit not for the one-particle density),  take the form of viscid Burgers equations, and  are valid for any $N$. This  procedure allows, in principle,  the calculation of all the
correlation functions for finite $N$ and $t$.

For the considered ensemble, the relevant set of orthogonal polynomials is that of Hermite polynomials, defined as~\cite{FYODOROVREV}  \beq h_k(x)=(-1)^k
e^{Nx^2/2} \frac{d^k}{dx^k} e^{-Nx^2/2}.  \eeq  These admit the following useful integral representation
\beq h_k(x)=(-iN)^k\sqrt{\frac{N}{2\pi}} e^{Nx^2/2}
\int_{-\infty}^{\infty} dq q^k e^{-N/2 q^2 +ixqN}. \label{integral}
 \eeq
 We shall also use the so-called monic polynomials, i.e. polynomials where the coefficient
 of highest order term is equal to unity:  $\pi_k(x) \equiv
 h_k(x)/N^k=\prod_{i=1}^k (x-\bar x_i)$, with $\bar x_i$ denoting the (real) zeros of the Hermite polynomials. In the random walk described above, the probability distribution retains its form at all instants of time. It is then easy to construct polynomials that remain orthogonal with respect to the time-dependent measure $\exp\left( -Nx^2/(2\tau)\right)$: all that needs to be done is to replace $N$ by $N/\tau$ in Eq.~(\ref{integral}). One obtains then
 \beq \pi_k(x,
\tau)=\frac{(-i)^k}{k!} \sqrt{\frac{N}{2 \pi \tau}} \!\int
dq q^k e^{-\frac{N}{2\tau} (q-ix)^2} ,\label{monictime} \eeq
which satisfy
\beq\label{normalpi}
\int_{-\infty}^{\infty} {dx}{\rm e}^{-\frac{N x^2}{2\tau}}\pi_n(x,\tau)\pi_m(x,\tau)=\delta_{nm}  c_n^2 ,
\eeq
with $c_n^2=n!\sqrt{2\pi}\tau^{n+1/2}$. The monic character of the $\pi_n$'s is not affected by the time dependence. By using the integral representation (\ref{monictime}), it is easy to show that the $\pi_n(x,\tau)$'s satisfy the following equation
 \beq\label{diffusionpi}
\partial_{\tau}  \pi_n(x,\tau)=-\nu_s\partial_{x}^2\pi_n(x,\tau),
\eeq
with $\nu_s= {1}/{2N}$. This is like a diffusion equation with however a negative
diffusion constant (note however that $\pi_n$ is an analytic function of $x$, and that $\pi_n(-iy,\tau)$, with $y$ real, satisfies a diffusion equation with a positive constant). At this
point, one may perform an inverse Cole-Hopf transform, i.e. , define the new function \beq \label{fkpoles} f_k(z,\tau)\equiv 2 \nu_s
\partial_z \ln \pi_k(z,\tau)=\frac{1}{N}\sum_{i=1}^k \frac{1}{z-\bar
x_i(\tau)}, \eeq with ${\rm Im}z\ne 0$. The resulting equation for $f_k$ is the
viscid Burgers~\cite{BURGERS} equation \beq\label{Burgersf}
\partial_{\tau}f_k(z, \tau)+ f_k(z,\tau)\partial_z f_k(z, \tau)=-\nu_s
\partial_{z}^2 f_k(z, \tau),
\eeq with $\nu_s$ playing the role of a  viscosity.

The equation (\ref{Burgersf}) is satisfied by all the functions
$f_k$. We shall focus now on the function $f_N$ associated to
$\pi_N(z,\tau)$. One reason is that $\pi_N(z,\tau)$ is known to be
equal to the average characteristic polynomial~\cite{BHPOLS}, i.e \beq \left <
\det (z-H(\tau)) \right> = \pi_N(z,\tau), \eeq
and, in the large $N$ limit, $(1/N)\del_z \ln  \langle\det (z-H(\tau)) \rangle\approx (1/N)\del_z \langle \ln \det (z-H(\tau)) \rangle=G(z)$.
Thus $f_N(z,\tau) $ coincides with the average resolvent $G(z,\tau)$ in
the large $N$ limit. In fact the structure of $f_N$, as clear from Eq.~(\ref{fkpoles}), is very close to that of the resolvent, with its poles given by the zeros of the characteristic polynomial.
Eq.~(\ref{Burgersf}) for   $f_N(z,\tau)$ is exact. The initial condition, $f_N(z,\tau=0)=1/z$, does not depend on $N$, so that all the finite $N$ corrections are taken into account by the viscous term. Note however that this exact equation does not allow us to directly study the finite $N$ corrections to the resolvent or the spectral density.

We turn now to the study of  the viscid Burgers equation for $f_N(z,t)$, i.e. Eq,~(\ref{Burgersf}) for $k=N$,  in the
vicinity of the (moving) pre-shock, that is for $x$ near
$z_c(\tau)$. The Cole-Hopf transformation used above provides a solution in term of $\pi_N(x,\tau)$ which allows us to study the effects of a small viscosity by performing a saddle point approximation on the integral (\ref{monictime}). The saddle point equation coincides with the characteristic equation discussed earlier (with the identification $\xi\to -iq$). However this analysis breaks down in the vicinity of the shock: here the difficulties comes from the merging of  the two saddle points associated with the square root singularity. A better analysis is then called for, and we shall rely in what follows on  tools borrowed from the theory of
turbulence~\cite{HOWLS}. Let us recall that in the vicinity of the
edge of the spectrum, and in the inviscid limit, \beq
f_N(z,\tau)\simeq
\pm\frac{1}{\sqrt{\tau}}\mp\frac{1}{\tau^{3/4}}\sqrt{z-z_c}. \eeq We
set \beq x=z_c(\tau)+\nu_s^{2/3}s, \quad  f_N(x,\tau)=\dot z_c(\tau)
+\nu_s^{1/3} \chi_N (s,\tau), \eeq with $\dot z_c\equiv
\partial_{\tau} z_c=\pm 1/\sqrt{\tau}$. The particular scaling of
the coordinate is motivated by the fact that near the square root
singularity the spacing between the eigenvalues scales as
$N^{-2/3}$. A simple calculation then yields the following equation
for $\chi(s,\tau)$ in the vicinity of $z_c(\tau)=2\sqrt{\tau}$: \beq
\del_\tau^2 z_c+\nu_s^{1/3}\frac{\del \chi}{\del
\tau}+\chi\frac{\del \chi}{\del s}=-\frac{\del^2 \chi}{\del s^2},
\eeq which, ignoring the small term of order $\nu_s^{1/3}$, we can
write as \beq \frac{\del}{\del s}\left[    -\frac{s}{2\tau^{3/2}}+
\frac{1}{2}\chi^2+\frac{\del \chi}{\del s}  \right]=0, \eeq or,
setting $\chi=2\del_s\ln \phi(s)$, as \beq \del_s^2
\phi-a(s-s_0)\phi=0,\quad a\equiv  \frac{1}{4\tau^{3/2}}, \eeq with
$s_0(\tau)$ an arbitrary function of $\tau$. The  general solution of this
equation is given by Airy functions \cite{Vallee}  $\phi_i(s)=\epsilon_i Ai(\epsilon_i(s-s_0)a^{1/3})$, where $\epsilon_i$ for $i=0,1,2$ are cubic roots of 1 and
$\sum_0^2 \phi_i(s)=0$. In the case of the characteristic polynomial, the solution corresponds to $\epsilon_0=1$.
 Finally, \beq
\chi(s,\tau)= 2\frac{Ai'(a^{1/3}(s-s_0))}{Ai(a^{1/3}(s-s_0))}. \eeq

We complete our hydrodynamic description of the Dyson fluid  by
considering the Cauchy transforms of the monic orthogonal polynomials
\beq \label{cauchy} p_k(z, \tau)=\frac{1}{2\pi i}\int_{-\infty}^{\infty}dx
\frac{ \pi_k(x,\tau)e^{-Nx^2/2\tau}}{x-z}. \eeq
The motivation for doing so is that  the
average of the inverse characteristic polynomial is related to
 $p_{N-1}$ ~\cite{SF}:  \beq \left< \frac{1}{\det (z-
H(\tau))}\right>=-\frac{2 \pi i}{c_{N-1}^2} p_{N-1}(z,\tau), \eeq
where $c_{N-1}^2$ is given after Eq.~(\ref{normalpi}).
 To get the time dependent version of the $p_k$'s,  we plug into Eq.~(\ref{cauchy})  the
 integral representation of the time-dependent monic polynomials $\pi_x(x,\tau)$. A simple calculation then yields, for
${\rm Im}z>0$, \beq p_k( z)=(-i)^k \sqrt{\frac{N}{2\pi
\tau}}\int_0^\infty  dq q^k e^{-\frac{N}{2}\tau (q^2-2i qz)} , \eeq
and $p_k(-z,\tau)=(-1)^{k+1}p_k(z,\tau)$. It can be shown by a direct
calculation that the functions \beq \tilde
p_k(z,\tau)=e^{\frac{N}{2\tau}z^2}p_k(z,\tau) \eeq satisfy the same
recurrence relations and differential equations as the polynomials
$\pi_k(z,\tau)$, in particular Eq.~(\ref{diffusionpi}). One may also verify that the function
  \beq g_k(z,\tau)=2 \nu_s\partial_z \ln
\tilde p_k(z,\tau),
\eeq
analogous to $f_k(z,\tau)$ in Eq.~(\ref{fkpoles}), satisfies a viscid Burgers equation:
 \beq
\partial_{\tau} g_k +g_k\partial_z g_k=-\nu_s
\partial _z^2g_k. \eeq
Also this  equation exhibits the phenomenon of universal pre-shock, but this time the scaling solution involves the two other solutions of the Airy
equation that were mentioned earlier, namely $\phi_1$ or $\phi_2$, depending on the sign of the imaginary part  of $z$, in agreement with~\cite{AF}.

 We note  that since  $\pi_N(z)$ and $p_N(z)$, and their first derivatives, are the building blocks  of all
relevant multipoint correlators (products of characteristic
polynomials, products of inverse characteristic polynomials and
mixed products~\cite{SF}),   all universal kernels originate implicitly
from  the  dynamics of shocks in the viscid Dyson fluid. Also, the so-called ``hard-edge'' scaling (Bessel kernels) can be be viewed in the same way, as a pre-shock approaching a hard wall.
We note also that other forms of universal shocks may appear, e.g.,  when the spectral density develops a cubic root singularity; then the  universal character of the phenomenon is coded in a Pearcey
function. This happens e.g. when two intervals supporting the
hermitian spectrum merge as a function of an external parameter
\cite{BREZINHIKAMI,NJPZ}, or when two Airy
type shock waves merge on a compact support ~\cite{NN,BN1}.

We expect the fluid dynamical picture presented here to hold also for other matrix-valued diffusion
processes, in particular for the
multiplicative diffusions, corresponding to  random walks involving products of complex matrices~\cite{GJJN,LNW}.
Perhaps the best studied example is the  random walk of unitary
matrices~\cite{JW}, because of its relation with two dimensional  Yang-Mills
theory with a large number of colors $N_c$~\cite{GOPAKUMAR}, and the associated universality conjectured by
Narayanan and Neuberger~\cite{NN}. We have earlier emphasized the relevance of shock waves
in this context~\cite{BN1}, as providing a simple mechanism for the
Durhuus-Olesen~\cite{DO} order-disorder transition.  Recently,   Neuberger  used the explicit expressions of averages of the characteristic polynomial and its inverse, that were obtained through a character expansion, to prove that closely related functions  satisfy Burgers equations with a  spectral viscosity $1/2N_c$ ~\cite{N1,N2}. The present analysis  supports the general character of these results and may motivate further generalizations. In
particular, we expect that pre-shock  phenomena may be identified  for
non-compact random matrix models, e.g. in mesoscopic systems where
the role of the time is played by the length of the
wire~\cite{BENAKKER}. Last but not least, this  picture of
shock formation finds  analogies in other branches of physics,
e.g., in the descritption of merging singularities  in  optics, leading to the so-called diffraction catastrophes (see e.g.~\cite{BERRY}),  or in the study of growth processes of the
Kardar-Parisi-Zhang universality class  and statistical properties
of the equilibrium shapes of  crystals~\cite{SPOHN}. We believe that pushing Dyson's original concept of temporal dynamics for random matrices  may allow for a better understanding  of above  analogies.

\section*{Acknowledgements}
We would like to thank Romuald A. Janik and Roland Speicher for illuminating remarks.
MAN thanks IPhT for hospitality during the time this paper was completed.  This work was supported by
  Marie Curie TOK Grant  MTKD-CT-2004-517186 ``Correlations in Complex Systems" (COCOS).



\end{document}